\documentclass{article}
\usepackage{ijcai25}
\usepackage{amssymb}
\usepackage{times}
\usepackage{color, soul}
\usepackage{url}
\usepackage[hidelinks]{hyperref}
\usepackage[utf8]{inputenc}
\usepackage[small]{caption}
\usepackage{graphicx}
\usepackage{xcolor}
\usepackage{amsmath}
\usepackage{amsthm}
\usepackage{booktabs}
\usepackage{algorithm}
\usepackage{algorithmic}
\usepackage[switch]{lineno}
\usepackage{multirow}
\usepackage{makecell}
\usepackage{colortbl}
\usepackage{subcaption}

\definecolor{darkgreen}{rgb}{0.0, 0.5, 0.0}
\definecolor{bluet5}{RGB}{100, 149, 237}   
\definecolor{greent5}{RGB}{60, 179, 113}   
\definecolor{yellowt5}{RGB}{255, 204, 102} 
\newcommand{\method}{\texttt{MPSL}}

\makeatletter
\renewcommand\subsubsection[1]{%
  \par\noindent\textbf{#1.}%
}
\makeatother

\title{
Fine-tuning Multimodal Transformers on Edge: \\
A Parallel Split Learning Approach
}

\author{
    Timo Fudala$^1$ \and Vasileios Tsouvalas$^1$\And Nirvana Meratnia$^1$
    \affiliations $^1$Eindhoven University of Technology\\
    \emails t.fudala@student.tue.nl, \{v.tsouvalas, n.meratnia\}@tue.nl
}

\begin{document}

\maketitle


\begin{abstract}
Multimodal transformers integrate diverse data types like images, audio, and text, advancing tasks such as audio-visual understanding and image-text retrieval; yet their high parameterization limits deployment on resource-constrained edge devices. Split Learning (SL), which partitions models at a designated cut-layer to offload compute-intensive operations to the server, offers a promising approach for distributed training of multimodal transformers, though its application remains underexplored. We present~\method\footnote{Code is available on~\href{https://github.com/Nousphera/MPSL}{https://github.com/Nousphera/MPSL}}, a parallel SL approach for computational efficient fine-tuning of multimodal transformers in a distributed manner, while eliminating label sharing, client synchronization, and per-client sub-model management.~\method~employs lightweight client-side tokenizers and a unified modality-agnostic encoder, allowing flexible adaptation to task-specific needs. Our evaluation across $7$ multimodal datasets demonstrates that~\method~matches or outperforms Federated Learning, reduces client-side computations by $250\times$, and achieves superior scalability in communication cost with model growth. Through extensive analysis, we highlight task suitability, trade-offs, and scenarios where~\method~excels, inspiring further exploration.
\end{abstract}

\section{Introduction} \label{introduction}

Multimodal transformers have demonstrated remarkable potential by integrating diverse data types such as images, audio, and text, thereby significantly advancing tasks like image captioning, audio-visual understanding, and vision-language question answering. The rapid proliferation of IoT devices, expected to surpass 29 billion by 2030~\cite{Bonaventura_2024}, generates a massive volume of data at the edge, offering transformative opportunities for AI applications across sectors such as healthcare, transportation, and communication. However, effectively leveraging this data necessitates addressing privacy concerns, regulatory restrictions, and the prohibitive cost of transferring vast datasets to centralized servers, which strains network bandwidth. Federated Learning (FL)~\cite{fl} has emerged as a promising solution by enabling collaborative model training without sharing raw data, allowing models to train locally while aggregating updates across participating clients. Despite its benefits, FL falls short in application to multimodal transformers due to their architectural complexity and overparameterization, making them unsuitable for training solely on resource-constrained edge devices~\cite{10605435}. 


To address these limitations, Split Learning (SL)~\cite{sl} has emerged as a compelling alternative, reducing client-side computation by partitioning a model and distributing it between edge devices and servers. SL enables edge devices to offload the computationally intensive portions of training to the server while preserving data privacy by keeping raw data on-device. Here, edge devices process initial layers locally and send intermediate activations — referred to as smashed data — to the server, which handles the remaining computations. Parallel Split Learning (PSL)~\cite{privacy-sensitive-psl} extends SL by enabling multiple devices to train simultaneously, addressing the latency issues of SL and making it a viable alternative to FL for computationally intensive models. However, while PSL has been studied for unimodal applications, its potential for handling multimodal data remains underexplored.

To bridge this gap, our work extends PSL to multimodal transformers, termed Multimodal Parallel Split Learning (\method), enabling computation-efficient training on edge devices. Inspired by~\cite{sasl}, we utilize a server-side loss aggregation mechanism that processes clients’ smashed data with a single backward pass, significantly reducing server-side computation and training latency compared to PSL. Additionally, we analyze the applicability of multimodal downstream tasks for PSL, identifying those that can be effectively trained and the factors influencing their performance. Our evaluation, which spans client-side communication overhead, computation overhead, and model performance, provides actionable insights into scenarios where PSL outperforms other distributed machine learning approaches in a multimodal setting. Concisely, our contributions are:

\begin{figure*}[!th]
    \centering \small
    \includegraphics[width=0.7\textwidth]{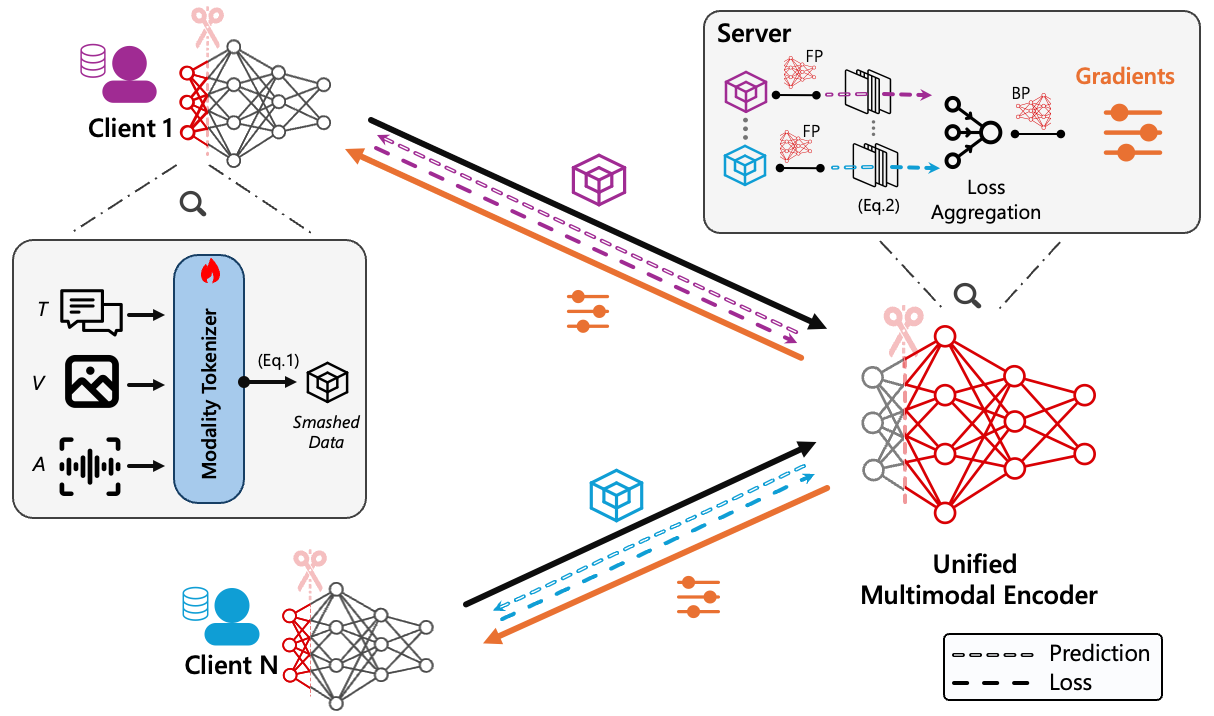}
    \caption{\small{An illustration of \method's training pipeline. Clients tokenize inputs of multiple modalities and offload intensive computations to the server, which processes the received activations through a unified encoder and then performs a single backward pass using an aggregated loss function. V: Vision, A: Audio, T: Text, BP: Back Propagation, FP: Forward Pass\label{fig:overview}}}
    \vspace{-10pt}
\end{figure*}

\begin{itemize}
\item We introduce~\method, extending PSL to enable computation-efficient fine-tuning of multimodal transformers on edge devices, eliminating label sharing, client-side model synchronization, and per-client sub-model management on the server.
\item ~\method~reduces client-side computational overhead by $250\times$ compared to FL regimes, while achieving performance comparable to centralized fine-tuning.
\item Our evaluation across $7$ datasets and multiple transformer sizes highlights task suitability, trade-offs, and scenarios where~\method~outperforms traditional distributed learning approaches in a multimodal setting. 
\end{itemize}

\section{Related Work} \label{related_work}

\noindent \textbf{Split Learning.} Split Learning (SL) was introduced to address the limitations of FL, particularly the high computational demands on resource-constrained edge devices. By partitioning the model at a designated cut-layer, SL enables edge devices to process initial layers locally and send intermediate activations (smashed data) to a server for forward and backward propagation. Although this approach alleviates client-side computational burdens, SL requires sharing labels; impeding label privacy~\cite{sl}. To address this, variations like U-shaped SL were proposed to ensure that both input features and labels remain on the client-side by partitioning the model into three segments: client-head, server-body, and client-tail \cite{sl,Shuffled-Transformer,MaskSL}. Despite these advancements, the sequential nature of vanilla and U-shaped SL imposed scalability limitations, prompting the development of Parallel Split Learning (PSL). 

Early PSL methods like LocalSplitFed~\cite{LocalSplitFed} required synchronizing client-side models, which added communication overhead, and maintaining a separate sub-model for each client on the server to enable parallelism~\textendash~a setup that becomes impractical with many clients or large models.
Subsequent PSL approaches overcome these limitations through several optimizations, such as using gradient aggregation~\cite{sglr,Mix2SFL,EPSL} to address client decoupling issues (as observed in SGLR~\cite{sglr}), and employing single-pass loss aggregation~\cite{sasl} to reduce redundant computation and improve scalability. Such advancements position PSL as a viable alternative for the distributed training of computationally intensive models. Despite this, existing PSL frameworks primarily focus on unimodal tasks, leaving the integration of multimodal data largely unexplored. This gap underscores the pressing need to advance PSL toward effective multimodal learning, where multimodal challenges such as modality alignment are further complicated by the decentralized nature of distributed training.

\noindent \textbf{Multimodal Transformers.} The emergence of multimodal transformers has enabled effective integration of diverse modalities such as images, audio, and text. Existing architectures typically follow two design paradigms: using modality-specific encoders or a unified encoder. Models like CLIP~\cite{clip} use separate encoders for each modality, aligning embeddings into a shared space via contrastive learning and demonstrating strong zero-shot transfer capabilities, particularly for image-text tasks. ImageBind~\cite{imagebind} extends this approach to additional modalities using modality-specific tokenizers and encoders, enabling generalization to unseen modality pairs through pretraining on image-paired data. ViT-LENS~\cite{ViT-LENS} adopts a more lightweight strategy by attaching modality-specific adapters to a frozen ViT encoder~\cite{ViT}, reducing data requirements while maintaining representational alignment. Despite their effectiveness, these models rely on modality-specific components or large backbone architectures, resulting in high parameter counts (e.g., up to $1.2$B for ImageBind) and architectural complexity~\textendash~factors that limit their suitability for distributed or resource-constrained environments. Unified encoder approaches have emerged as a more scalable alternative, aiming to share parameters across modalities. VATT~\cite{vatt} employs a modality-agnostic encoder with shared weights to process video, audio, and text, reducing model size compared to modality-specific designs. Meta-Transformer~\cite{meta-transformer} extends this idea by supporting $12$ modalities through a single encoder and lightweight modality-specific tokenizers, demonstrating strong performance in parameter-efficient multimodal learning. Nonetheless, such models are predominantly designed for centralized training and remain computationally intensive for on-device learning.

Recently, parameter-efficient fine-tuning (PEFT) approaches have been explored for distributed training of multimodal transformers, with methods like FedCLIP~\cite{FedCLIP} applying lightweight adapters on top of a frozen backbone within a FL framework. However, FL still requires clients to execute the frozen encoder backbone, limiting its applicability to resource-constrained edge devices. PSL offers a promising alternative, yet its application to multimodal transformers remains largely underexplored.~\method~ extends PSL to multimodal transformers, addressing client-side computational efficiency and modality alignment, while also eliminating the need for label sharing.

\section{\texttt{MPSL} Framework} \label{methodology}
We present a Multimodal Parallel Split Learning (\method) framework for fine-tuning multimodal transformers, eliminating the need for label sharing, synchronization of client-side models, and maintaining per-client sub-models on the server. Specifically,~\method~extends~\cite{sasl} to support multiple modalities, enabling support for multimodal transformer architectures. By offloading the computationally intensive components of multimodal transformers training to the server-side,~\method~significantly reduces the computational overhead on clients. Furthermore, we utilize an aggregated client-side loss to perform a single, unified backward pass on the server.

\subsection{Notations} \label{notations}
We use the following notations in the remainder of the paper. 
Let $\mathcal{N}=\{n_1, ..., n_N\}$ be the set of all clients, with $N$ as the total number of clients. For each client $n \in \mathcal{N}$, let $\mathcal{D}_n$ be its local dataset and $\mathcal{B}_n \subset \mathcal{D}_n$ a mini-batch of $\mathcal{D}_n$. The global batch is defined as $\mathcal{B}=\cup_{n \in \mathcal{N}} \mathcal{B}_n$. 
For a set of input modalities $\mathcal{M}=\{m_1, ..., m_M\}$, let $T_m$ be a modality-specific tokenizer for each modality $m \in \mathcal{M}$. The multimodal transformer model, denoted as $W$, has weight parameters $\theta = (\theta_1, \dots, \theta_d) \in \mathbb{R}^d$  and is divided into three components: the head ($W_h$, modality-specific tokenizers), the body ($W_b$, encoder), and the tail ($W_t$, task-specific classifier). Hence, $W = [W_h;W_b;W_t]$, where $W_h$ consists of a set of tokenizers with $W_h=\{T_m\}_{m \in \mathcal{M}}$. 

We split $W$ into a server-side model $\mathcal{F}_S = [W_b;W_t]$ and client-side model $\mathcal{F}_C = W_h$, with $\mathcal{F}_{C_n}$ as the client model for a client $n$. For a client $n$ with $M$ modalities, a single data sample is defined as $z=(\{\mathbf{x}_{m_i}\}_i^M, y) \in \mathcal{B}_n$, where $\{\mathbf{x}_{m_i}\}_i^M = (\mathbf{x}_{m_1}, \dots, \mathbf{x}_{m_M})$ represents the multimodal input, and $y$ denotes the label (i.e., the ground truth). The activations computed by $\mathcal{F}_{C_n}$ for a given input $\{\mathbf{x}_{m_i}\}_i^M$ are denoted as $\mathbf{a}_n$, while the predictions with respect to these activations computed by the server model $\mathcal{F}_S$ are expressed as $\hat{y} = \mathcal{F}_S(\mathbf{a}_n)$. Lastly, we define $\mathcal{L}_{C_n}$ as the $n$-th client's loss function, and with $\mathcal{L}_S$ we denote the aggregated server-side loss function, given by $\mathcal{L}_S = \sum_{n \in \mathcal{N}} {\frac{|\mathcal{B}_n|}{|\mathcal{B}|} \mathcal{L}_{C_n}}$.

\subsection{Modality fusion in~\method} \label{method:multimodal_fusion}
Leveraging a unified encoder supports both early and late fusion of modalities. Inspired by~\cite{meta-transformer}, we use modality-specific tokenizers to tokenize inputs and fuse modalities by concatenating their representations into a single joint vector. In early fusion, tokenized modalities are concatenated into a joint vector on the client-side and processed jointly by the server-side encoder. In late fusion, each modality is tokenized and processed independently by the server-side encoder, with their encoded representations concatenated afterward into a single joint vector. Regardless of the fusion type, we apply global average pooling to the multimodal joint vector to obtain the final multimodal embedding, which is then used for the model’s final prediction. \\

\noindent \textbf{Processing multimodal inputs on clients.} For a given client $n$ with a data sample $z=(\{\mathbf{x}_{m_i}\}_i^M, y) \in \mathcal{B}_n$, the client extracts the activations $\mathbf{a}_n$ as follows: 
\vspace{-5pt}
\begin{equation}\label{formula:client_activations}
    \resizebox{0.4\textwidth}{!}{
    $
    \mathbf{a}_n=\mathcal{F}_{C_n}(\{\mathbf{x}_{m_i}\}_i^M) =
    \begin{cases} 
        [T_{m_i}(\mathbf{x}_{m_i})]_i^M ~~\text{if } \text{fusion} = early \\
        {T_{m_i}(\mathbf{x}_{m_i})}_i^M ~~~~~~~\text{otherwise}
    \end{cases}
    $
    }
\end{equation}

\noindent where $T_{m_i}$ is the modality-specific tokenizer for modality $i$. The activations $\mathbf{a}_n$ are subsequently sent to the server, which in turn computes the prediction $\hat{y}$ with respect to $\mathbf{a}_n$, and transmits back to the client. Given the label $y$ for $\{\mathbf{x}_{m_i}\}_i^M$, the client-side loss is computed locally on the client-side as $\mathcal{L}_{C_n}(\hat{y}, y)$ and sent to the server to compute the gradients. Since client $n$ only sends its loss to the server, label privacy is preserved. Lastly, using the received cut-layer gradients from the server, client $n$ then performs backpropagation on its model $\mathcal{F}_{C_n}$. Formally, the client-side objective function is:

\begin{equation}\label{eq:client_opt}
    \underset{\mathcal{F}_{C_n}}{\text{min}} \quad \mathcal{L}_{C_n}(\hat{y}, y) 
\end{equation}


\noindent \textbf{Server-side training.} We further elaborate on the training process on the server-side. Upon receiving the activations $\mathbf{a}_n$ from a client $n$, the server computes its corresponding prediction $\hat{y}$ as follows:
\vspace{-5pt}
\begin{equation}\label{formula:server_prediction}
    \resizebox{0.4\textwidth}{!}{
    $
    \hat{y}=\mathcal{F}_S(\mathbf{a}_n) = 
    \begin{cases}
        W_t(\mathrm{GAP}({W_b(\mathbf{a}_n))}) ~~\text{if } \text{fusion} = early \\
        W_t(\mathrm{GAP}([W_b(\mathbf{a}_{n_{m_i}})]_1^M)) ~\text{otherwise}
    \end{cases}
    $
    }
\end{equation}

\noindent where $\mathrm{GAP}$ denotes the global average pooling operator.
The server subsequently sends prediction $\hat{y}$ to client $n$. Note that in the case of late fusion, each modality in $\mathbf{a}_n$ is encoded independently using $W_b$, concatenated, and then processed with $\mathrm{GAP}$, ultimately leading to the final prediction $\hat{y}$ via $W_t$.

The server computes the global aggregated loss $\mathcal{L}_S$ after receiving $\mathcal{L}_{C_n}$ from all clients $n \in \mathcal{N}$ and performs a single backward pass on $\mathcal{F}_S$. This approach reduces server-side computational burden by requiring only one backward pass on the computationally intensive encoder $W_b$, regardless of number of clients, and eliminates the need to maintain multiple client sub-models. It is important to note that~\method~requires no explicit synchronization between client-side models. Instead, coordination emerges implicitly through shared cut-layer gradients computed from the aggregated global loss. This design mitigates client drift by aligning local updates around a common global signal and removes the need to manage individual client models on the server-side. After backpropagation, the server sends the cut-layer gradients to all clients $n \in \mathcal{N}$. Formally, the server-side objective function is:
\vspace{-5pt}
\begin{equation}
    \underset{\mathcal{F}_S}{\text{min}} \quad \mathcal{L}_S = \sum_{n \in \mathcal{N}} {\frac{|\mathcal{B}_n|}{|\mathcal{B}|} \mathcal{L}_{C_n}}
\end{equation}

\subsection{Post-training Model Construction} \label{method:final_model}
After training, the splitted multimodal transformer can be reassembled without restrictions. We impose no limitations in this regard; one approach is to construct a complete model for each client $n$ by concatenating the client-side model $\mathcal{F}_{C_n}$ with the server-side model $\mathcal{F}_S$, forming [$\mathcal{F}_{C_n}; \mathcal{F}_S$]. This approach is suitable for personalization. Alternatively, a single client-side model $\mathcal{F}_{C_{\text{agg}}}$ can be obtained by aggregating all client-side models, following FedAvg~\cite{fl}, which is concatenated with the server-side model $\mathcal{F}_S$, resulting in [$\mathcal{F}_{C_{\text{agg}}}$;$\mathcal{F}_S$]. We adopt the latter approach to directly compare~\method's performance with other FL regimes~\cite{fl,FedCLIP}. 

\section{Experiments} \label{experiments}

\noindent \textbf{Datasets.} We evaluate all combinations of image, audio, and text modality pairs across diverse downstream tasks and datasets. From (image, text) pairs, we use T4SA~\cite{T4SA} (B-T4SA subset) for sentiment analysis, COCO-QA~\cite{COCO-QA} for visual question answering, and MS-COCO~\cite{COCO} and Flickr30K~\cite{Flickr30K} for image-text retrieval. For (image, audio) pairs, we leverage Kinetics-Sounds~\cite{kinetics-sound} and UCF101~\cite{UCF101} for action recognition. Finally, (audio, text) pairs from MELD~\cite{MELD} are used for multimodal emotion classification. For all data splitting procedures, we use a Dirichlet distribution over classes, denoted as $\mathrm{Dir}$($0.1$), following~\cite{li2021federatedlearningnoniiddata}. \\

\noindent \textbf{Models.} Throughout our experiments, we use pre-trained weights from Meta-Transformer~\cite{meta-transformer} based on ViT-B/16~\cite{ViT}, fine-tuning the complete encoder for image-text retrieval and the last $6$ encoder blocks for classification tasks, unless stated otherwise. In Figure~\ref{fig:ablation_2__2__scaling_encoder}, where we use various ViT encoder depths (Tiny, Small, Large, and Huge), we fine-tune the latter half of the encoder blocks using pre-trained weights from~\cite{ViT}. Late fusion was employed in all cases except for COCO-QA, T4SA, and MELD, where early fusion yields the best performance (see Table~\ref{table:ablation_3}). For image-text retrieval we use the contrastive loss from~\cite{ONE_PEACE}, while for all classification tasks we employ Cross-Entropy loss. \\

\noindent \textbf{Modality Tokenization Details.} To obtain uniform representations, we use modality-specific tokenizers similar to~\cite{meta-transformer,imagebind}. Images are ``\textit{patchified}" following ViT~\cite{ViT}, text is tokenized similar to CLIP~\cite{clip}, and raw audio is converted to spectrograms and subsequently ``\textit{patchified}" following AST~\cite{AST}. Additionally, for audio and/or vision modalities, a  $\mathrm{cls}$ (classification) token was prepended to the tokenized representation to capture a global representation of the entire input modality, allowing the concatenation of only the $\mathrm{cls}$ token(s) instead of the complete embedding vector during late fusion. \\
    
\noindent \textbf{Baselines.} We evaluate~\method~against centralized fine-tuning and two distributed learning baselines: standard FedAvg~\cite{fl}~and FedCLIP~\cite{FedCLIP}. As noted in Section~\ref{related_work}, prior PSL work for distributed multimodal settings is scarce, motivating the development of~\method. Since no existing PSL methods apply in this context, we focus our comparison on representative FL baselines: FedAvg, as a foundational method, and FedCLIP, which incorporates PEFT~\cite{PEFT} to enable efficient client-side adaptation. While other FL variants exist, they typically still depend on local model training followed by parameter aggregation and synchronization. We therefore compare~\method~to the most relevant and practical alternatives for distributed multimodal learning. We evaluate performance across $3$ criteria: (1) task-specific performance (e.g., accuracy for classification, recall for retrieval), (2) client-side communication overhead (average MB transmitted per client per epoch), and (3) computational overhead (measured in G-FLOPs per input and the number of trainable client-side parameters). All distributed baselines are trained for the same number of rounds to ensure a fair comparison. To reduce the effect of randomness during training, we report the mean test accuracy over $3$ runs with different random seeds.

\begin{table*}[!t]
    \centering \small
    \resizebox{0.85\textwidth}{!}{%
    \begin{tabular}{clcccccccc}
        \toprule
        & \multirow{2}{*}{\textbf{Method}} 
        & \multicolumn{2}{c}{\cellcolor{blue!5}\includegraphics[height=0.55cm]{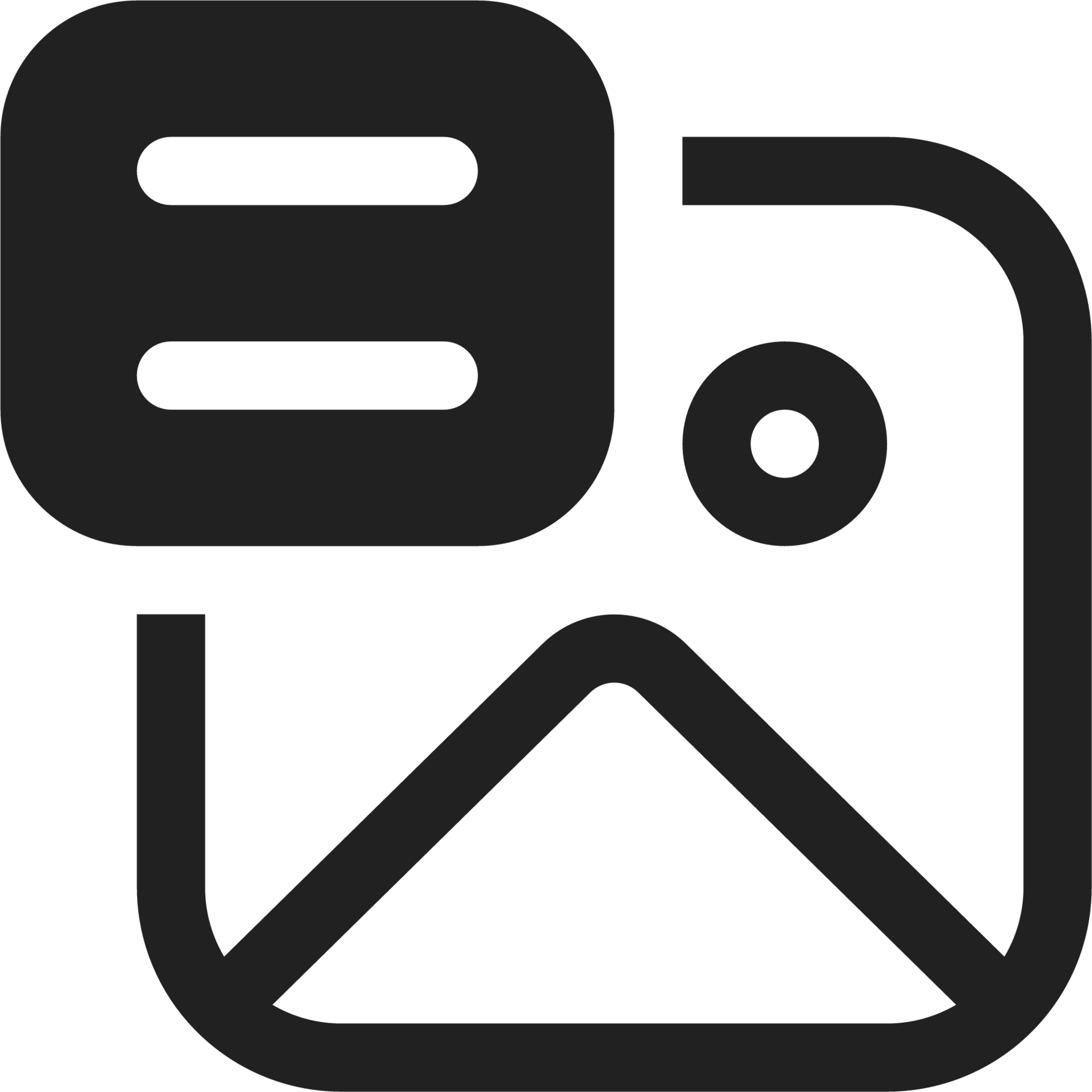} \textbf{(V+T)}}
        & \multicolumn{2}{c}{\cellcolor{green!5}\includegraphics[height=0.55cm]{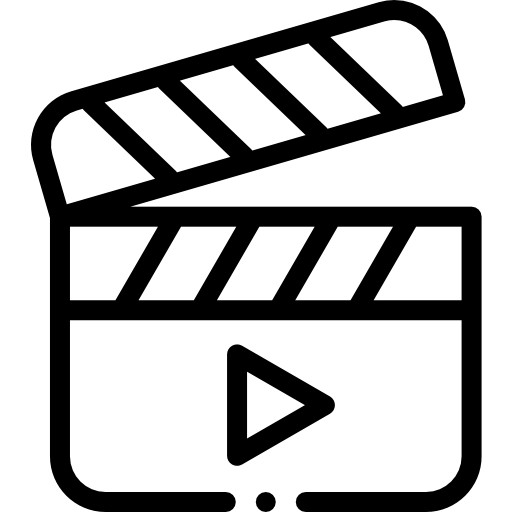} \textbf{(V+A)}}
        & \cellcolor{yellow!5}\makecell{\includegraphics[height=0.55cm]{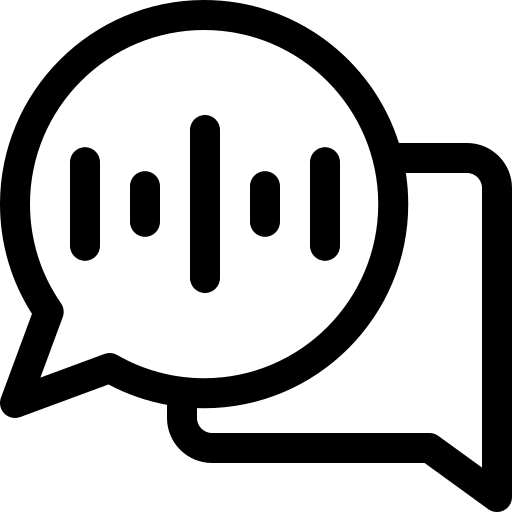} \textbf{(A+T)} \vspace{2pt}}
        & \multirow{2}{*}{\makecell{\textbf{Avg. Tr.}\\ \textbf{Params} (\textbf{M})}}
        & \multirow{2}{*}{\makecell{\textbf{Avg. Comp.}\\(\textbf{GFlops})}} \\ 
        \cmidrule(lr){3-4} \cmidrule(lr){5-6} \cmidrule(lr){7-7}

        & & \textcolor{bluet5}{\textbf{COCO-QA}} & \textcolor{bluet5}{\textbf{T4SA}} 
        & \textcolor{greent5}{\textbf{Kinetics-Sounds}} & \textcolor{greent5}{\textbf{UCF101}} 
        & \textcolor{yellowt5}{\textbf{MELD}} & & & \\ 
        \midrule

        \multicolumn{2}{c}{\textbf{Centralized}} & 61.7 $\pm$ 0.1 & 83.0 $\pm$ 0.1 & 81.8 $\pm$ 0.2 & 91.3 $\pm$ 0.2 & 61.4 $\pm$ 0.4 & 43.7 & 25.8 \\ \cmidrule(lr){1-10}
        \multirow{3}{*}{\textbf{N=25}} 
            & \textbf{FedAvg}         & \textbf{59.1 $\pm$ 0.2} & \textbf{76.2 $\pm$ 0.6} & \textbf{78.2 $\pm$ 0.3} & \textbf{90.8 $\pm$ 0.6} & 57.4 $\pm$ 0.6 & 43.7 & 25.8 \\
            & \textbf{FedCLIP}        & 32.0 $\pm$ 0.2 & 55.8 $\pm$ 5.0 & 70.1 $\pm$ 0.4 & 70.0 $\pm$ 3.9 & 55.3 $\pm$ 0.2 & 3.1 & 25.8 \\
            & \textbf{MPSL (\textit{Ours})}     & 53.3 $\pm$ 0.4 & 73.3 $\pm$ 0.0 & 72.2 $\pm$ 1.6 & 90.4 $\pm$ 0.5 & \textbf{59.4 $\pm$ 0.5} & \textbf{1.0} & \textbf{0.1} \\

        \midrule

        \multirow{3}{*}{\textbf{N=100}} & 
            \textbf{FedAvg}         & \textbf{53.5 $\pm$ 0.2} & 72.4 $\pm$ 1.6 & \textbf{71.4 $\pm$ 0.4} & \textbf{89.8 $\pm$ 0.9} & 48.9 $\pm$ 0.0 & 43.7 & 25.8 \\
            & \textbf{FedCLIP}        & 13.1 $\pm$ 2.6 & 45.0 $\pm$ 2.8 & 63.5 $\pm$ 0.8 & 56.4 $\pm$ 1.2 & 50.7 $\pm$ 0.7 & 3.1 & 25.8 \\
            & \textbf{MPSL (\textit{Ours})}     & 46.2 $\pm$ 0.5 & \textbf{72.5 $\pm$ 1.2} & 67.0 $\pm$ 1.1 & 89.3 $\pm$ 1.0 & \textbf{58.2 $\pm$ 0.9} & \textbf{1.0} & \textbf{0.1} \\

        \bottomrule
    \end{tabular}%
    }
    \caption{Performance evaluation of classification tasks. Computation overhead is reported as the average number of trainable parameters (in millions) and GFlops on the client-side. V: Vision, A: Audio, T: Text.}
    \label{tab:main_res}
\end{table*}

\begin{table*}[!t]
    \centering \small
    \resizebox{0.77\textwidth}{!}{%
    \begin{tabular}{clccccccc}
        \toprule
        & \multirow{2}{*}{\textbf{Method}}
        & \multicolumn{2}{c}{\textcolor{bluet5}{\textbf{MS-COCO}}}
        & \multicolumn{2}{c}{\textcolor{bluet5}{\textbf{Flickr30K}}}
        & \multirow{2}{*}{\makecell{\textbf{Avg. Tr.}\\\textbf{Params} (\textbf{M})}}
        & \multirow{2}{*}{\makecell{\textbf{Avg. Comp.}\\(\textbf{GFlops})}} \\ 
        \cmidrule(lr){3-4} \cmidrule(lr){5-6}
        & & \textbf{\textit{R@1}} & \textbf{\textit{R@10}} & \textbf{\textit{R@1}} & \textbf{\textit{R@10}} & & & \\
        \midrule
        \multirow{4}{*}{\textbf{\textit{Image-to-Text}}} 
            & \textbf{Centralized} & 56.8 $\pm$ 2.4 & 77.8 $\pm$ 1.9 & 66.8 $\pm$ 5.3 & 86.0 $\pm$ 2.9 & 85.8 & 8.65 \\ \cmidrule(lr){2-9}
            & \textbf{FedAvg} & 1.5 $\pm$ 0.1 & 5.2 $\pm$ 0.3 & 11.2 $\pm$ 1.7 & 28.0 $\pm$ 3.4 & 85.8 & 8.65 \\
            & \textbf{FedCLIP} & 0.1 $\pm$ 0.0 & 0.3 $\pm$ 0.1 & 0.1 $\pm$ 0.0 & 0.9 $\pm$ 0.3 & 2.52 & 8.65 \\
            & \textbf{MPSL (\textit{Ours})} & \textbf{21.1 $\pm$ 0.8} & \textbf{43.5 $\pm$ 1.2} & \textbf{26.0 $\pm$ 0.5} & \textbf{49.2 $\pm$ 1.2} & \textbf{0.74} & \textbf{0.1} \\
        \midrule
        \multirow{4}{*}{\textbf{\textit{Text-to-Image}}}
            & \textbf{Centralized} & 54.3 $\pm$ 2.1 & 92.1 $\pm$ 1.0 & 62.4 $\pm$ 5.2 & 95.2 $\pm$ 1.7 & 85.8 & 8.65 \\ \cmidrule(lr){2-9}
            & \textbf{FedAvg} & 1.5 $\pm$ 0.1 & 10.4 $\pm$ 0.4 & 9.3 $\pm$ 1.2 & 42.9 $\pm$ 3.6 & 85.8 & 8.65 \\
            & \textbf{FedCLIP} & 0.1 $\pm$ 0.0 & 0.9 $\pm$ 0.2 & 0.2 $\pm$ 0.0 & 1.4 $\pm$ 0.1 & 2.52 & 8.65 \\
            & \textbf{MPSL (\textit{Ours})} & \textbf{17.3 $\pm$ 0.2} & \textbf{60.3 $\pm$ 1.2} & \textbf{21.6 $\pm$ 0.9} & \textbf{65.5 $\pm$ 1.1} & \textbf{0.74} & \textbf{0.1} \\
        \bottomrule
    \end{tabular}%
    }
    \vspace{-5pt}
    \caption{Performance evaluation of retrieval tasks with 100 clients. Computation overhead is reported as the average number of trainable parameters (in millions) and GFlops on the client-side.}
    \label{tab:retrieval_res}
\end{table*}


\subsection{\method~Evaluation}

\noindent \textbf{Model Performance.} We evaluate~\method~on $7$ datasets and compare its performance against two distributed baselines in Tables~\ref{tab:main_res} and~\ref{tab:retrieval_res}. As shown in Table~\ref{tab:main_res}, in classification tasks,~\method~consistently matches FedAvg across all modality pairs and tasks while significantly reducing client-side computation. By offloading the computationally intensive model components to the server,~\method~achieves over a $250$-fold reduction in GFLOPs, requiring only $1.0$M trainable parameters~\textendash~a 97.71\% decrease compared to FedAvg. Notably,~\method~outperforms FedAvg in certain settings, such as MELD, demonstrating its potential for both efficiency and performance. In contrast, for image-text retrieval (Table~\ref{tab:retrieval_res}), all $3$ distributed approaches exhibit a notable performance drop compared to centralized training. Despite this,~\method~surpasses both FL baselines while maintaining client-side efficiency on par with classification tasks. We further elaborate on the gap between distributed and centralized performance for retrieval in Section~\ref{ablation_mini_batch_size}. \\

\begin{figure}[!t]
    \vspace{-5pt}
    \centering \small
    \includegraphics[width=0.8\linewidth]{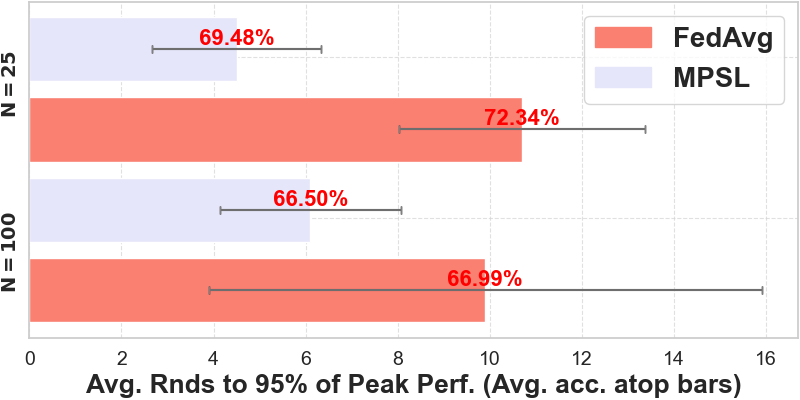}
    \caption{\small{Average number of rounds to reach 95\% of peak performance. Numbers above bars denote average final accuracy.}}
    \label{fig:convergence_analysis}
    \vspace{-10pt}
\end{figure}

\noindent \textbf{Convergence Analysis.} We compare the convergence speed of~\method~and FedAvg (excluding FedCLIP due to its poor performance; see Table~\ref{tab:main_res}) by measuring the number of rounds to reach 95\% of peak accuracy. As shown in Figure~\ref{fig:convergence_analysis},~\method~converges significantly faster, especially in the $N{=}100$ regime where data is highly distributed. This stems from their differing update strategies: FedAvg relies on local backpropagation and global averaging, which can lead to client drift, while~\method~uses a single server-side backpropagation over aggregated losses, providing a stabilizing global signal. This enables~\method~to converge more quickly and remain stable even in distributed multimodal settings, where modality misalignment compounds the effects of data heterogeneity and further hinders optimization. \\

\begin{figure}[!t]
    \vspace{-25pt}
    \centering \small
    \includegraphics[width=0.8\linewidth]{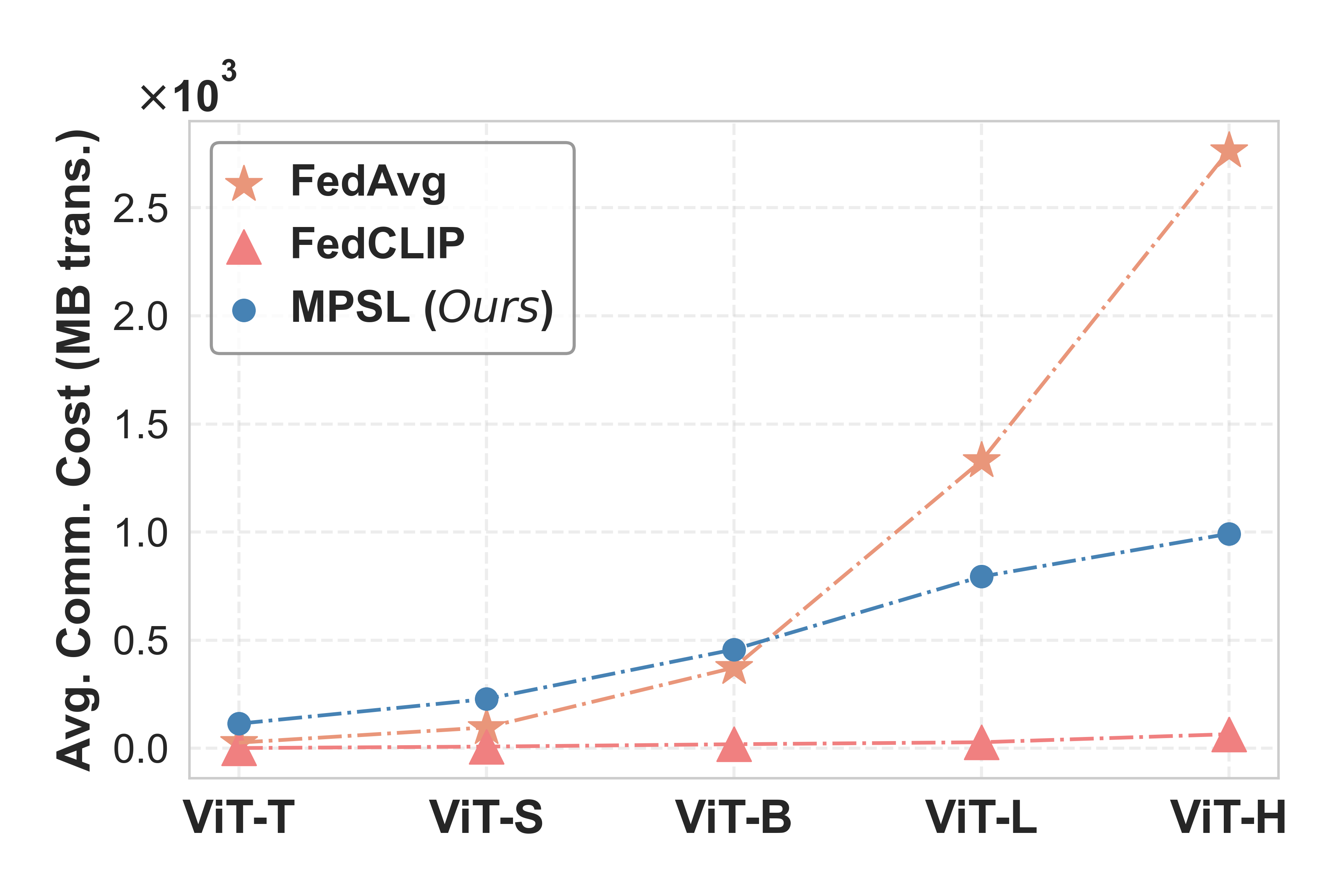}
    \vspace{-10pt}
    \caption{\small{Impact of encoder depth on client-side communication overhead compared to FedAvg.}}
    \label{fig:main_results__comm_overhead_w_scaling}
    \vspace{-5pt}
\end{figure}

\noindent \textbf{Communication Cost.} Unlike FL regimes that depend on transmitting trainable parameters,~\method~only communicates smashed data (i.e., activations), decoupling communication from model size. As shown in Figure~\ref{fig:main_results__comm_overhead_w_scaling}, FedAvg can be more communication-efficient for small ViT models (e.g., Tiny, Small), but this benefit fades with larger encoders — common in multimodal transformers~\cite{imagebind}. FedCLIP, using lightweight adapters on a frozen backbone, achieves the lowest communication cost overall; however, its performance remains limited (see Table~\ref{tab:main_res}). In contrast,~\method~consistently outperforms FedCLIP across all tasks while maintaining lower computation. This highlights~\method's suitability for scalable multimodal learning under strict communication constraints.

\subsection{\method~Ablations}\label{ablation}
Here, we delve into the design choices underlying~\method~and rigorously assess their impact on diverse downstream tasks. This analysis sheds light on the applicability of downstream tasks for multimodal SL, pinpoints those that can be effectively trained using~\method, and uncovers the critical factors that underpin their performance. \\

\subsubsection{Batch size effect\label{ablation_mini_batch_size}} Batch size plays a critical role in modality alignment for multimodal learning, where smaller sizes can cause feature collapse in the unified embedding space~\cite{clip,duan2022multimodalalignmentusingrepresentation}. To examine this in SL settings, we evaluate the impact of batch size in~\method~across downstream tasks (classification and image-text retrieval) and modality pairs. By keeping all SL settings fixed and varying the batch size, we investigate whether similar effects occur, with results presented in Table~\ref{table:ablation_1__result_with_25_clients}. From Table~\ref{table:ablation_1__result_with_25_clients}, it can be observed that batch size significantly impacts model performance, with sensitivity varying by task. For classification tasks like COCO-QA and UCF101, increasing the batch size from $50$ to $500$ improves accuracy by approximately 15\%, while T4SA shows a more modest gain of $6\%$. 

\begin{figure}[t!]
    \centering
    \begin{subfigure}[b]{0.495\linewidth}
        \centering \small
        \includegraphics[width=\linewidth, trim=100 100 100 100, clip]{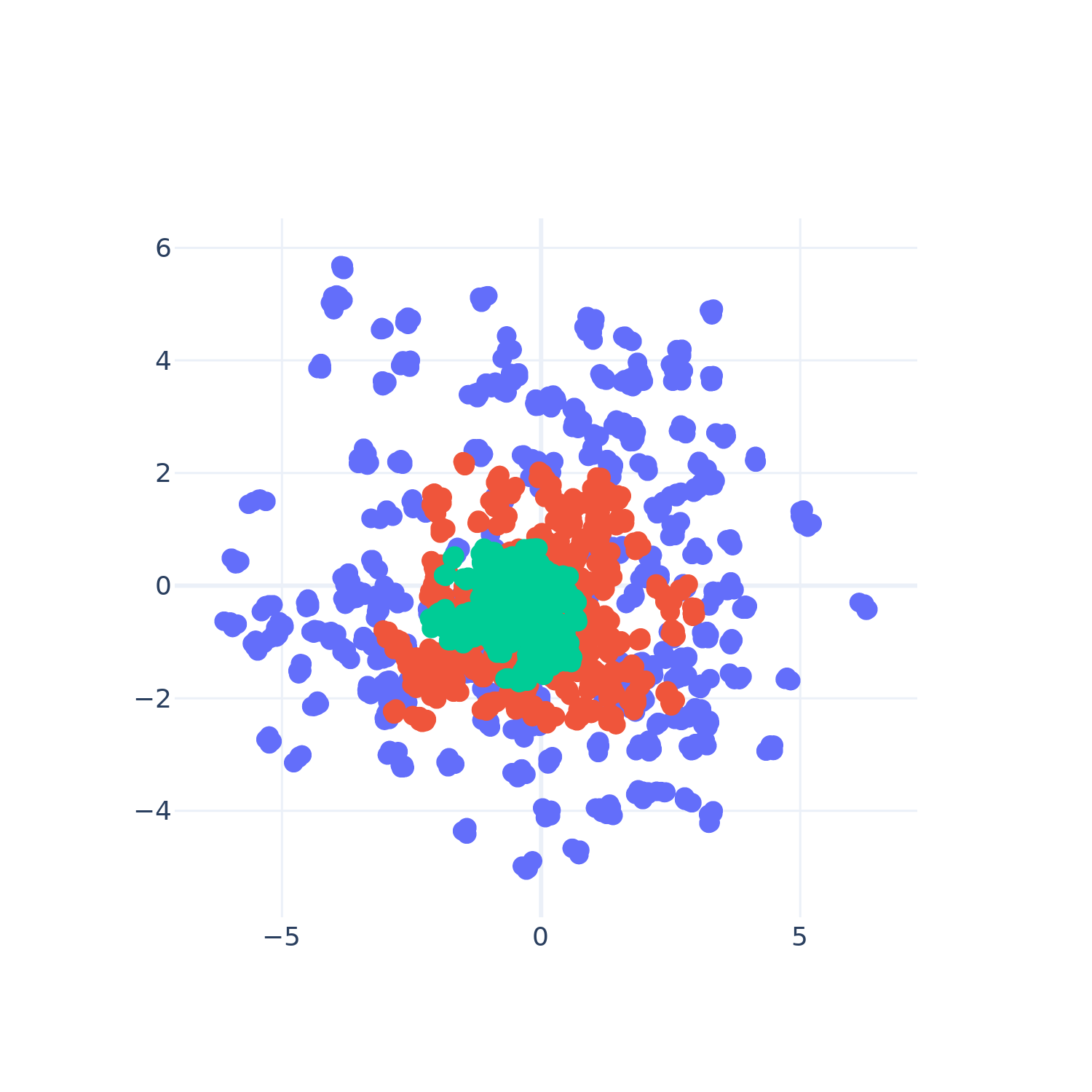}
        \caption{\small{MS-COCO - \textit{Text}}} \label{ablation_1__subfig:a}
    \end{subfigure}
    \vspace{-2pt}
    \begin{subfigure}[b]{0.495\linewidth}
        \centering
        \includegraphics[width=\linewidth, trim=100 100 100 100, clip]{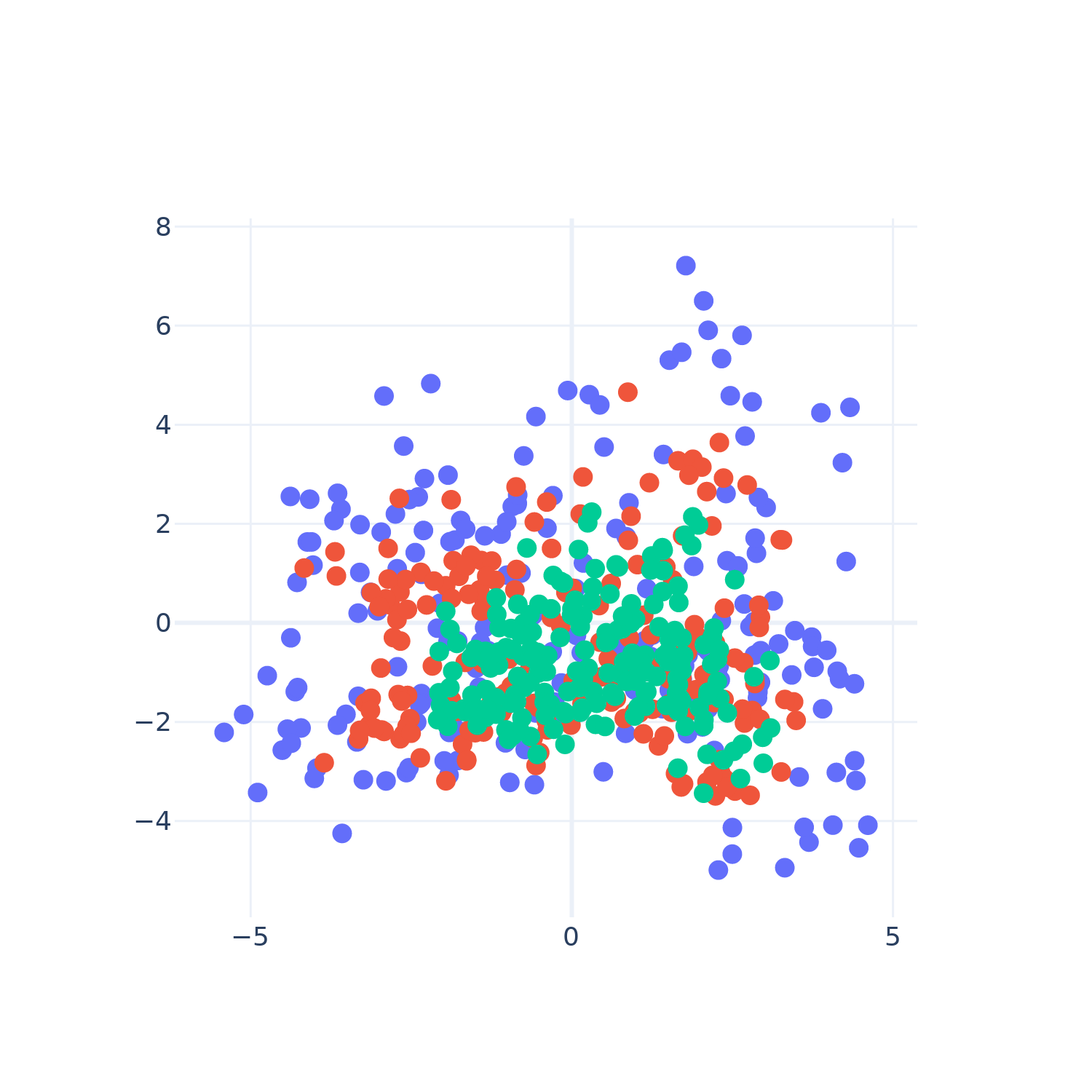}
        \caption{\small{MS-COCO - \textit{Image}}} \label{ablation_1__subfig:c}
    \end{subfigure}
    \vspace{-5pt}
    \begin{subfigure}[b]{0.495\linewidth}
        \centering
        \includegraphics[width=\linewidth, trim=100 100 100 100, clip]{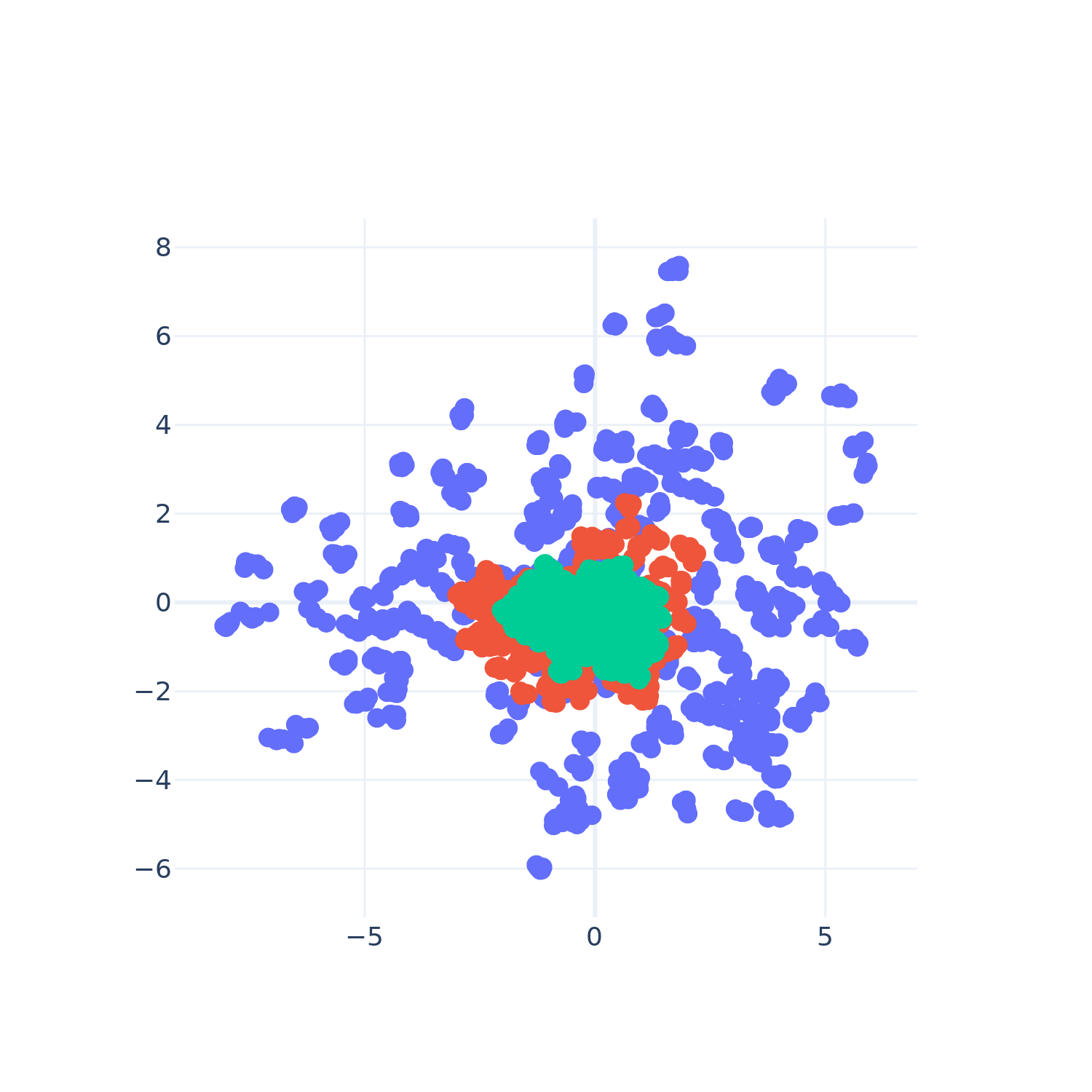}
        \caption{\small{Flickr30K - \textit{Text}}} \label{ablation_1__subfig:b}
    \end{subfigure}
    \vspace{-2pt}
    \begin{subfigure}[b]{0.495\linewidth}
        \centering
        \includegraphics[width=\linewidth, trim=100 100 100 100, clip]{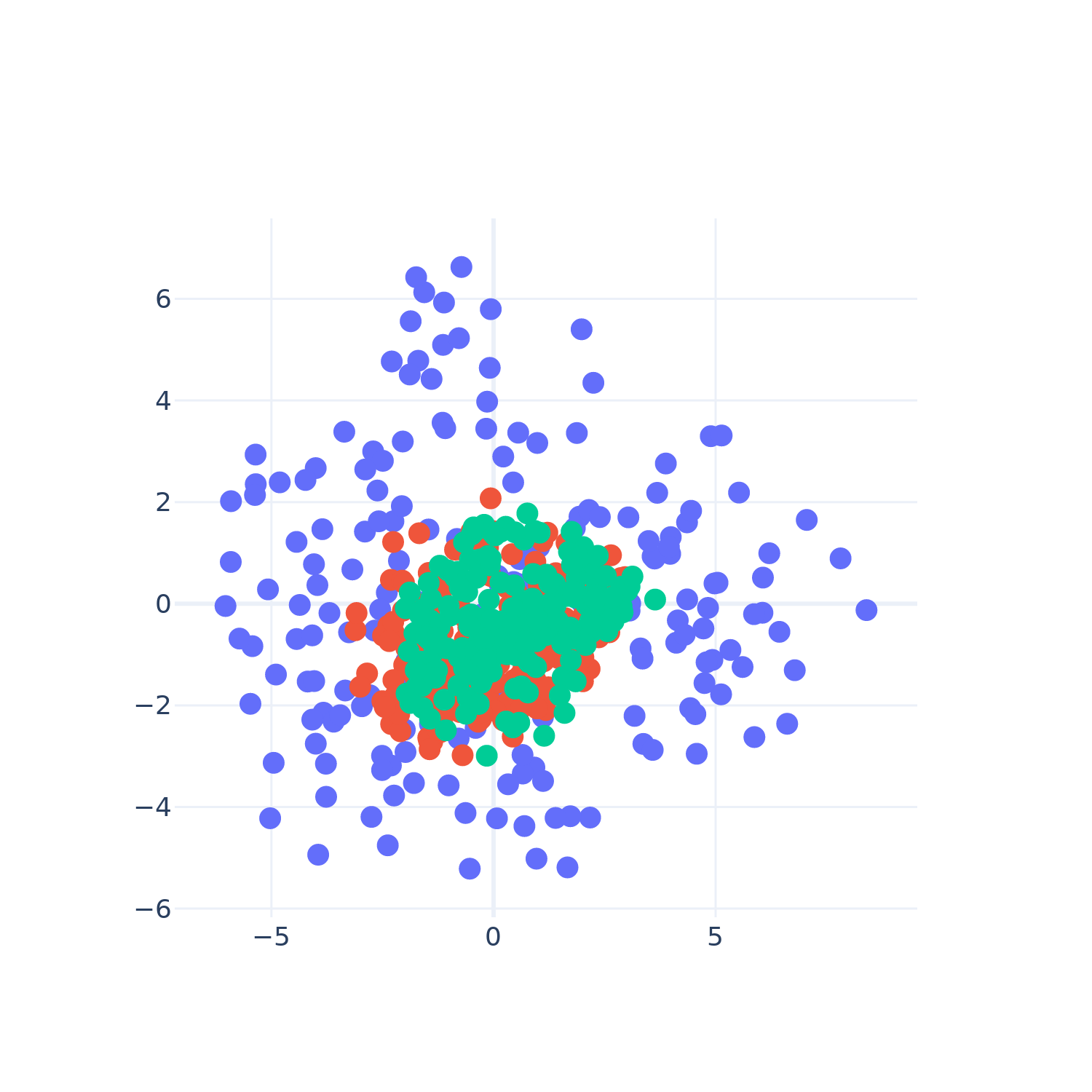}
        \caption{\small{Flickr30K - \textit{Image}}} \label{ablation_1__subfig:d}
    \end{subfigure}
    \caption{\small{Effect of batch size on embedding spaces after PCA for Flickr30K and MS-COCO with $N$=$25$. \textcolor{blue}{Blue} represents embeddings from centralized models, while \textcolor{darkgreen}{green} and \textcolor{red}{red} indicate~\method’s embeddings with batch size of $50$ and $200$, respectively.}}
    \label{fig:ablation_1__embedding_visualization}
    \vspace{-5pt}
\end{figure}

In contrast, image-text retrieval tasks show the most substantial improvements, with recall increasing by $\approx$40\% on average as batch sizes grow from $50$ to $150$ and $200$, underscoring the importance of larger batch sizes for retrieval tasks. Intuitively, these tasks depend on maximizing the distance between unrelated embeddings and minimizing it for related ones; thus, proper alignment is essential to distinguish relevant embeddings from irrelevant ones in the shared embedding space. Larger batch sizes improve this alignment by introducing diverse examples within each batch, reducing feature collapse and significantly enhancing retrieval performance. We further depict this in Figure~\ref{fig:ablation_1__embedding_visualization}, where we visualize the embedding space for image and text modalities at batch sizes of $50$ and $200$, along with centralized fine-tuning. Larger batch sizes result in less dense clusters, indicating better alignment through increased separation among dissimilar embeddings. However, larger batch sizes require clients to hold more samples, which makes tasks where edge devices hold a limited number of labeled samples less practical for SL settings. Sequential techniques to simulate large batch sizes could alleviate this limitation, but they fall outside the scope of this work. \\

\begin{table*}[t!]
    \centering  \small
    \resizebox{0.6\linewidth}{!}{%
        \begin{tabular}{lccccccc}
            \toprule
            \textbf{Batch size} & 50 & 100 & 150 & 200  & 300 & 400 & 500 \\ 
            \midrule
            \textbf{COCO-QA}~\cite{COCO-QA} 
            & 69.4 & 73.6 & \textendash & 78.3 & 80.9 & 83.0 & \textbf{85.3} \\
            \textbf{T4SA}~\cite{T4SA} 
            & 83.4 & 87.8 & \textendash & 85.3 & 87.8 & \textbf{89.4} & 86.8 \\
            \textbf{UCF101}~\cite{UCF101}  
            & 84.5 & 95.4 & \textendash & 97.6 & \textbf{100.0} & \textbf{100.0} & 98.5 \\
            \midrule
            \textbf{Flickr30K}~\cite{Flickr30K} 
            & 37.9 & 69.5 & 63.0 & \textbf{69.6} & \textendash & \textendash & \textendash \\
            \textbf{MS-COCO}~\cite{COCO} 
            & 21.3 & 53.2 & 64.1 & \textbf{69.7} & \textendash & \textendash & \textendash \\ \bottomrule
        \end{tabular}%
    }
    \vspace{-5pt}
    \caption{\small{Impact of batch size on model performance across tasks with $N$=$25$. We report normalized accuracy and top-1 recall vs. centralized fine-tuning for classification and retrieval tasks, respectively. Entries with ``\textendash'' reflect infeasible data partitioning due to dataset size.}}
    \label{table:ablation_1__result_with_25_clients}
    \vspace{-10pt}
\end{table*}

\subsubsection{Trainable parameters\label{ablation_2}} While our experiments primarily use ViT-B, the multimodal transformers utilized in multimodal learning have substantial variations in encoder depths, spanning from millions~\cite{meta-transformer,vatt} to a billion parameters~\cite{imagebind}. This raises the question of how well our findings generalize to other, often larger, encoders. To address this, we conduct experiments by varying both the number of fine-tuned blocks within the ViT-B encoder and the total encoder depth. \\

\begin{figure}[!t]
    \centering \small
    \vspace{-10pt}
    \includegraphics[width=0.9\linewidth]{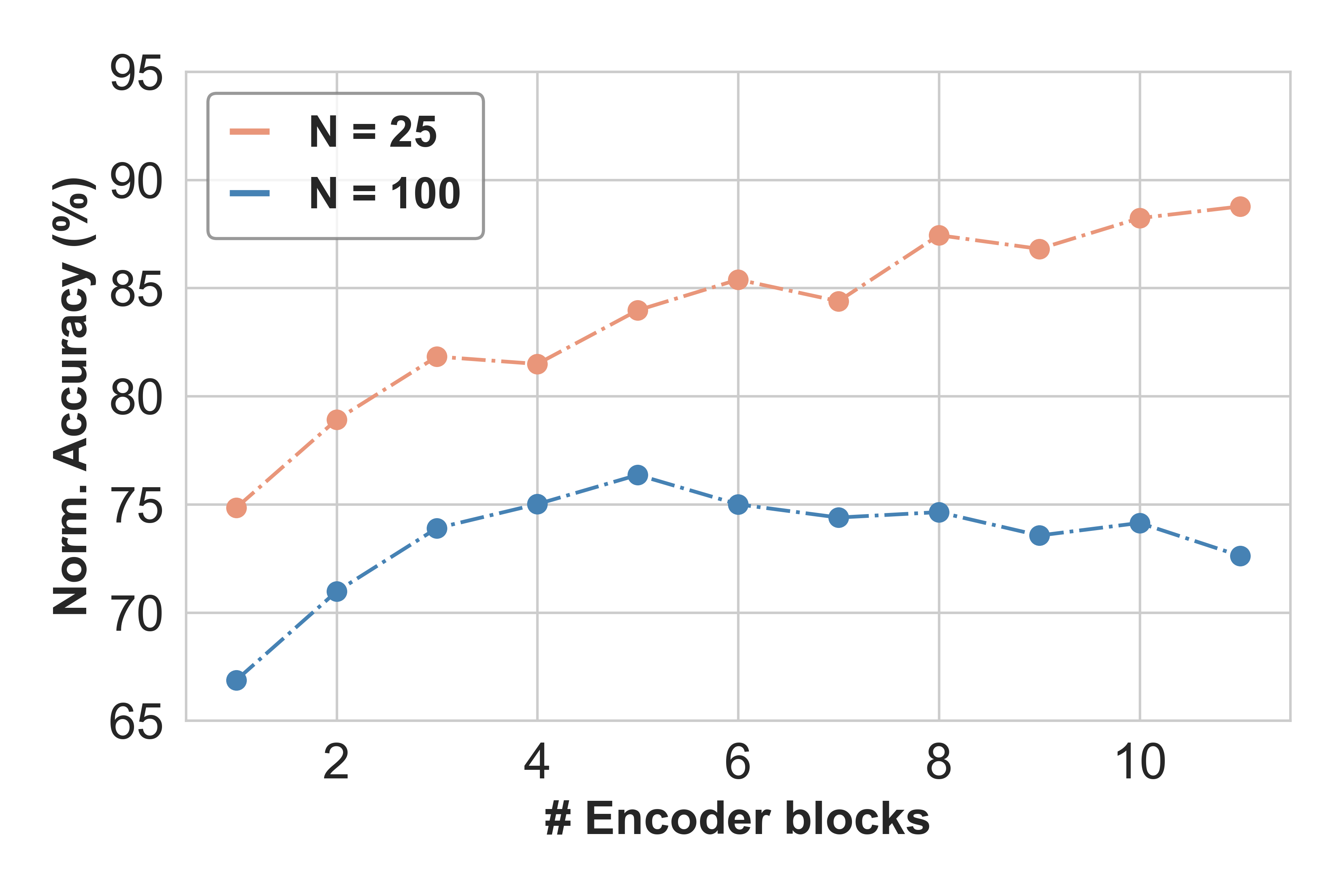}
    \vspace{-5pt}
    \caption{\small{Impact of fine-tuned ViT blocks on model performance for COCO-QA.}}\label{fig:ablation_2__1__coco_qa__all_blocks}
    \vspace{-5pt}
\end{figure}

\noindent \textit{Number of Fine-tuned Blocks.} We conduct experiments across all tasks, varying the number of trainable parameters by fine-tuning an increasing number of encoder blocks while keeping other SL settings fixed. Notably, this increase in trainable parameters is limited to the server-side, imposing no additional computational burden on clients. As shown in Table~\ref{table:ablation_2__1}, increasing the number of trainable parameters does not exhibit a consistent correlation with task-specific performance, as results vary across tasks. However, our findings indicate a minimum threshold of fine-tuned blocks is necessary to achieve sufficient performance, with fine-tuning a single block consistently resulting in poor performance. This is also evident in a more elaborate analysis in COCO-QA, presented in Figure~\ref{fig:ablation_2__1__coco_qa__all_blocks}, where performance improves sharply from 1 to 3 blocks but plateaus beyond that point. \\

\begin{figure}[t]
    \centering \small
    \vspace{-10pt}
    \includegraphics[width=0.9\linewidth]{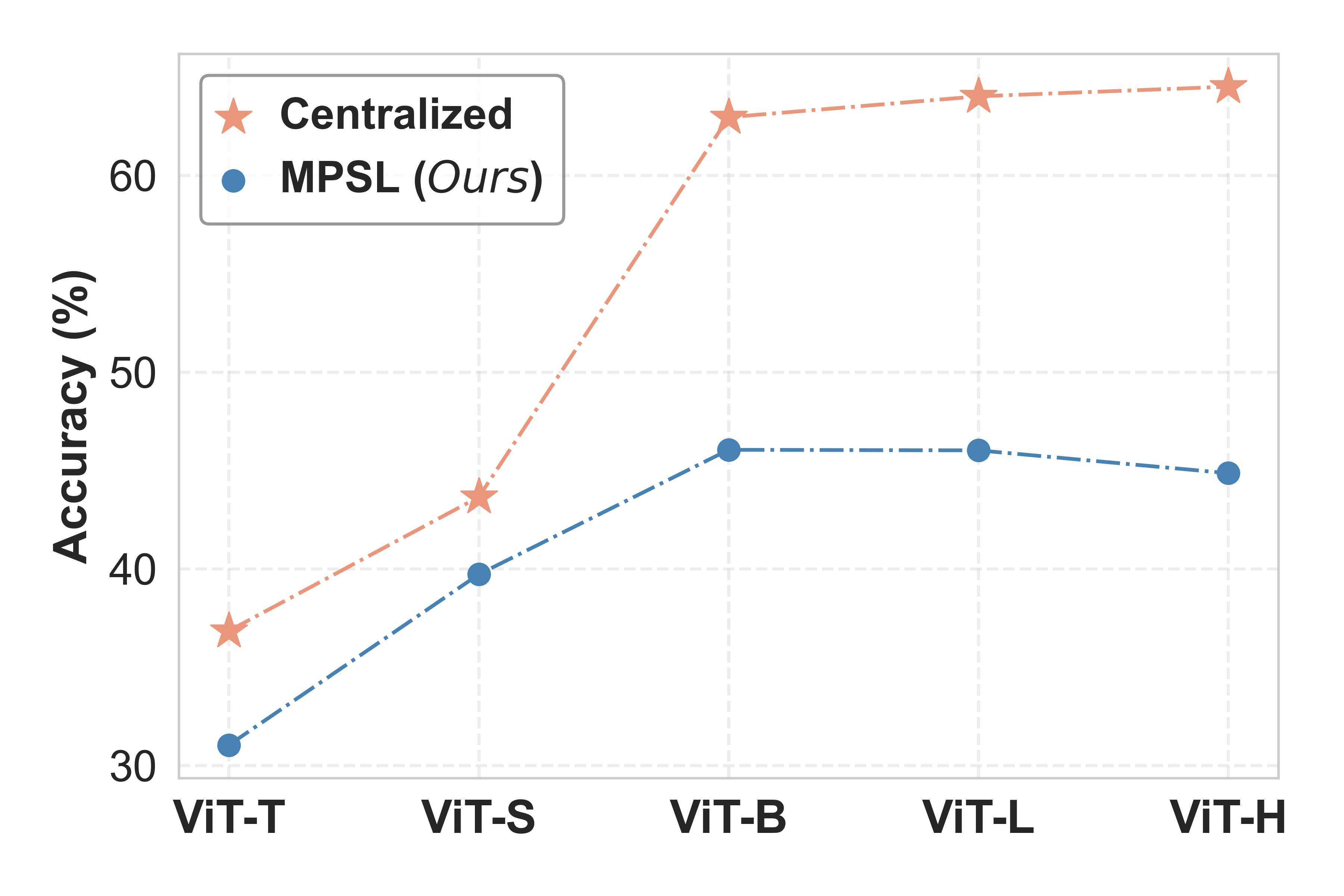}
    \vspace{-5pt}
    \caption{\small{Impact of \textit{encoder depth on performance} for COCO-QA.}} \label{fig:ablation_2__2__scaling_encoder}
    \vspace{-5pt}
\end{figure}

\begin{table}[t!]
    \centering \small
    \resizebox{0.97\linewidth}{!}{%
        \begin{tabular}{llllllll}
            \toprule
            \multirow{2}{*}{\textbf{\# ViT blocks}} & \multicolumn{3}{c}{\textbf{N $=$ 25}} & & \multicolumn{3}{c}{\textbf{N $=$ 100}} \\ 
            \cmidrule(lr){2-4} \cmidrule(lr){6-8}
             & \textit{~1} & \textit{~6} & \textit{~12} & & \textit{~1} & \textit{~6} & \textit{~12} \\
            \midrule
            \textbf{COCO-QA}~\cite{COCO-QA} & 74.8 & 85.4 & \textbf{88.8} &  & 66.7 & \textbf{74.5} & 70.4 \\
            \textbf{T4SA}~\cite{T4SA} & 80.7 & 88.5 & \textbf{90.9} &  & 75.7 & 89.0 & \textbf{89.6} \\
            \textbf{Kinetics-Sounds}~\cite{kinetics-sound} & 80.1 & 90.4 & \textbf{91.3} &  & 64.0 & \textbf{78.4} & 71.4 \\
            \textbf{UCF101}~\cite{UCF101} & 94.6 & 98.5 & \textbf{98.7} &  & 92.7 & 96.5 & \textbf{98.1} \\
            \textbf{MELD}~\cite{MELD} & 90.2 & 95.6 & \textbf{96.0} &  & 85.8 & 95.6 & \textbf{96.4} \\
            \midrule
            \textbf{Flickr30K}~\cite{Flickr30K} & \multicolumn{1}{c}{-} & \textbf{62.5} & 56.8 &  & \textendash & 36.1 & \textbf{36.4} \\
            \textbf{MS-COCO}~\cite{COCO} & 27.6 & 54.5 & \textbf{68.5} &  & 7.0  & 24.1 & \textbf{37.0} \\ 
            \bottomrule
        \end{tabular}%
    }
    \caption{\small{Performance across tasks when fine-tuning the last 1, 6, or all ViT blocks, with 25 and 100 clients. We report test accuracy and top-1 Image-to-Text recall (for Flickr30K and MS-COCO), normalized to centralized baselines.}}
    \label{table:ablation_2__1}
    \vspace{-5pt}
\end{table}

\noindent \textit{Encoder Depth.} We evaluate the impact of encoder depth on task-specific performance in the SL setting by scaling the encoder across all ViT~\cite{ViT} variations: ViT-Ti, ViT-S, ViT-B, ViT-L, and ViT-H, with 6, 22, 85, 303, and 630 million parameters, respectively. Based on the observations in Figure~\ref{fig:ablation_2__1__coco_qa__all_blocks}, we fine-tune the latter half of the encoder blocks for each variation: the last 6 blocks for ViT-Ti, ViT-S, and ViT-B, and the last 12 and 16 blocks for ViT-L and ViT-H, respectively, while keeping all other SL parameters fixed. Due to computational constraints with ViT-H, we use a fixed batch size of 100 across all variations, unlike prior experiments that optimize batch size for performance. This ensures consistent configurations during evaluation rather than achieving optimal performance for each model. Our findings in Figure~\ref{fig:ablation_2__2__scaling_encoder} reveal that SL performance remains stable even as model size significantly increases. For example, scaling from ViT-B (85 million parameters) to ViT-H (630 million parameters) increases trainable parameters from 43 million to 315 million, yet there is no notable drop in performance, highlighting~\method's scalability and robustness of larger models in SL settings. \\

\subsubsection{Fusion Type}\label{ablation_3} We examine the impact of fusion type (i.e., \textit{early vs. late}) on model performance across tasks, aiming to identify whether a task-agnostic modality fusion mechanism exists in SL settings. To this end, we conduct experiments with early and late fusion across all tasks requiring fusion (i.e., classification tasks) using $100$ clients ($N$=$100$) and report our findings in Table~\ref{table:ablation_3}.

We observe that the choice of fusion type significantly impacts model performance in the SL setting; however, no single fusion approach is superior across all tasks. Instead, the effectiveness of modality fusion is task-dependent. For instance, tasks involving vision-text (V+T) modalities generally perform better with early fusion, whereas vision-audio (V+A) tasks favor late fusion — patterns that hold consistently across both centralized and SL settings. Notably, in some cases, we observe a performance improvement exceeding 20\% when switching between fusion types. Conversely, for audio-text (A+T) modalities, the fusion choice has a less pronounced effect on performance. Thus, while fusion type undeniably impacts task performance, its effect is highly context-dependent, varying notably across tasks and modality pairs in the SL setting.

\begin{table}[t]
    \centering \small
    \resizebox{0.9\linewidth}{!}{%
        \begin{tabular}{clccccc}
        \toprule
        \multirow{2}{*}{\textbf{Modality}} & \multirow{2}{*}{\textbf{Dataset}} & \multicolumn{2}{c}{\textbf{Centralized}} &  & \multicolumn{2}{c}{\textbf{\method} (Ours)} \\ 
        \cmidrule(lr){3-4} \cmidrule(lr){6-7} 
        & & \textit{early} & \textit{late} & & \textit{early} & \textit{late} \\ 
        \midrule
        \textbf{V+T} & \textbf{COCO-QA}~\cite{COCO-QA} & \textbf{61.6} & 60.7 &  & \textbf{46.1} & 45.3 \\
        \textbf{V+T} & \textbf{T4SA}~\cite{T4SA} & \textbf{83.0} & 81.9 & & \textbf{62.4} & 45.2 \\
        \textbf{V+A} & \textbf{Kinetics-Sounds}~\cite{kinetics-sound} & 80.7 & \textbf{81.7} & & 49.9 & \textbf{66.2} \\
        \textbf{V+A} & \textbf{UCF101}~\cite{UCF101} & 90.2 & \textbf{90.5} &  & 86.5 & \textbf{88.1} \\ 
        \textbf{A+T} & \textbf{MELD}~\cite{MELD} & 62.1 & \textbf{62.5} &  & \textbf{58.9} & 56.4 \\
        \bottomrule
    \end{tabular}%
    }
    \caption{\small{Impact of \textit{early vs. late} modality fusion on performance across classification tasks. V: Vision, A: Audio, T: Text.}}
    \label{table:ablation_3}
    \vspace{-10pt}
\end{table}

\section{Conclusions}
We introduce \method, a Multimodal Parallel Split Learning framework that can be used to fine-tune multimodal transformers in a distributed manner with lightweight client computational burden. It uses an aggregated loss function of local client losses and a single unified backward pass on the server to eliminate the need for label sharing, synchronization of client-side models, and maintaining per-client sub-models on the server, alongside employing lightweight modality-specific tokenizers on the client-side to reduce computational overhead. Our experiments across multiple datasets, types of tasks, and pairs of modalities demonstrate \method's effectiveness in fine-tuning such models with comparable or superior performance to FL regimes while significantly decreasing client-side computational burden, and achieving superior scalability in communication cost with model growth. Future work involves addressing real-world challenges such as heterogeneous edge hardware, client dropouts, and missing modalities across clients.

\small{
\section*{Acknowledgments}
This work was funded by the DAIS project, which has received funding from ECSEL Joint Undertaking under grant agreement No 101007273, and the Bio-Curity project, which has received funding from Eureka Cluster Xecs under grant agreement No 2022016.
}

\clearpage
\bibliographystyle{named}
\bibliography{refs}

\end{document}